\documentstyle[12pt]{article}
\begin{document}
\begin{center}
{\Large{\bf Composite Fermions Phenomenology in Quantum Hall Effect}}
\vskip0.15cm
{\bf Keshav N. Shrivastava}
\vskip0.15cm
{\it School of Physics, University of Hyderabad,\\
Hyderabad  500046, India}
\end{center}
\vskip0.5cm
An effort is made to understand the phenomenological composite fermion model of the quantum Hall effect. The odd denominators are composed by adding  $\pm 1$ to the even numbers, 2, 4, 6, and 8. Although the denominators are phenomenologically made to agree with the experimental data, the physics of the CF model ,i.e., the {\sl even flux quanta attachment} to the electron is found to be incorrect.
\vskip0.1cm
\noindent keshav@mailaps.org   Fax. +91-40-3010145
\vskip1.0cm
\baselineskip24pt
\section{Introduction}

In 1980 von Klitzing et al [1] have shown that the Hall conductivity
is given by an integer multiple of $e^2/h$ so that an accurate
value of $e^2/h$ can be measured. Later on, it was found by Tsui
et al [2] that not only the integer but also the fractional
multiples of $e^2/h$ can be identified at high magnetic fields.
Recently, Pan et al[3] have obtained the measurements of the quantum Hall effect upto a magnetic field of 42 T. They have obtained  the effective charge of a fraction of unity from the minima in the longitudinal resistivity. The minima are found at $\nu$=2/11, 3/17
3/19, 2/13, 1/7, 2/15, 2/17 and 1/9. It is believed that magnetic flux is quantized so that the field can be written as,
$$
B= {nhc\over \nu e}\eqno(1)
$$
where $n$ is the electron density per unit area, $n=n_o/A$. We can define the effective charge of the quasiparticles as,
$$
e_{eff}=\nu e\eqno(2)
$$
so that $\nu$ determines the fraction which gives the charge. We can 
write,
$$
B= s\phi_o = {nhc\over \nu e}\eqno(3)
$$
so that
$$
s=n/\nu, \phi_o=hc/e.\eqno(4)
$$
So far we have used only one concept, i.e., flux is quantized except that the effective charge is taken to be different from $e$. We have
 not used the concept of Landau levels. We shall interpret the data of longitudinal resistivity of GaAs/AlGaAs. We define the resistivity as,
$$
\rho_{xx} ={h\over\nu e^2}.\eqno(5)
$$
Since the effective charge is $\nu e$, we can pick up the minima in $\rho_{xx}$ and measure the value of $\nu$ experimentally and tabulate them. Here the job of the experimentallist is finished. In this way
the charge is parameterised. It is another problem to find the theory which changes the charge $e$ to $\nu e$.

\section{CF phenomenological model}

     We discuss the composite fermion, (CF),  phenomenological model
first suggested by Jain[4]. We propose that there are two types of particles, the electrons and the CFs. They both see the quantized field except that the field seen by CF is not equal to that seen by the electron. Let us say that B$^*$ is seen by the CF and B by the electron. The quantization conditions are,
$$
\nu B = n\phi_o (electrons)\eqno(6)
$$
and
$$
\nu^*B^*=n\phi_o (CF).\eqno(7)
$$
Usually all particles should see the same value but in this picture different particles see different values. This is justified by finding
 a way to go from B to B$^*$. We consider the even number $2p$ of 
fluxes attached to the electron so that $2pn\phi_o$ is a field which measures the difference between the field seen by the electron and that seen by CF. The field can be parallel to that of the electron or it can be antiparallel so that we can use both the signs, $\pm$ and write the field experienced by the CF as,
$$
B^* = B\pm 2pn\phi_o.\eqno(8)
$$
Thus the field seen by the CF is given by the electron field and the flux field. Now substituting (6) and (7) in (8) gives,
$$
\nu^* ={\nu\over (1\pm 2p\nu)}.\eqno(9)
$$
For p = 1,
$$
\nu^* = {\nu\over(2\nu +1)}, {-\nu\over (2\nu - 1)}.\eqno(10)
$$
Let us flip the sign of $\phi_o$ and use other even numbers so that,
$$
|\nu^*|= {\nu\over(2\nu\pm 1)}, {\nu\over(4\nu\pm 1)}, {\nu\over(6\nu\pm 1)}, {\nu\over(8\nu\pm 1)}.\eqno(11)
$$
These series can be identified in the experimental data and it appears that identification of the series is equivalent to observation of the CF. This is the CF model and every thing appears to be fine.

\section{Composite Fermions and Biot and Savart's law}

     Are there any concepts in the CF phenomenological model? If the idea was to find the series of charges, there was no trouble at all. However, if the series is to be linked to some theory then there is trouble. From eq.(6) and eq.(7),
$$
\nu B= \nu^*B^*\eqno(12)
$$
The factor $\nu$ has the charge. Therefore, in order to keep the product constant, we can increase the charge and reduce the field so that $\nu B$ is unchanged. In the Biot and Savart's law, the field can be reduced by reducing the current or charge, i.e., the field {\sl cannot} be reduced by increasing the charge. On the otherhand in the CF, the field is reduced by increasing the charge. Since the current is flow of charge, the result (12) is not in conformity with the Biot and Savart's law. In eq(11), the sign of $\nu^*$ has been changed in one of the formulae. The equation (8) already has both signs. The question is whether the sign of $\phi_o$ can be changed in the flux quantization.
 If the field is to be written as a quantized flux, the sign of the field should be the same as that of the flux. Therefore, eqs.(6) and 
(7) do not allow to change the sign of the flux $\phi_o$, otherwise
 the flux $\phi_o$ will become equal to zero. Therefore, the use of 
both signs in (8) is not consistent. In spite of the problem with 
the sign of the flux and the Biot and Savart's law, the CF agrees 
with the data. When we start with the  fractions of charges which 
are the same as the experimental values, and work out the fields and fluxes, we find that there is disagreement with the established laws. Therefore the assumptions of the model are not correct.

     Let us consider the Biot and Savart's law. The flow of current, 
$I$ or charge, $e$, produces a magnetic field,$B$. This is usually achieved by making a coil so that the field produced by the current 
is along the axis of the coil. Since $2pn\phi_o$ is a field, the current, $I_{\phi}$, associated with this field is to be determined 
by $2pn\phi_o$=2$I_{\phi}/cR_1$ where $R_1$ is the radius of 
curvature of the coil and $c$ is the velocity of light. Now if
 there is a need to attach the field $2pn\phi_o$ to $B$, then the current $I_{\phi}$ must be added to $I$. Then we can get the field $B+2pn\phi_o$ but the charge corresponding to $I_{\phi}$ must also be carried to $e$. It is not possible to change the field without changing the charge. This means that the flux can be attached only when charges are also attached. The attachment of flux in (8) without any regard
 of adding currents is not allowed by the Biot and Savart's law. Attaching two flux quanta to one electron is thus not allowed. In
 Fig.1, we draw a picture of coil current and field. The current I
gives the field B as shown in Fig.1(a), I$_{\phi}$ gives 
B$_{\phi}$ as shown in Fig.1(b). The Fig.1(c) shows the attachment of two flux quanta to one electron and Fig.1(d) four flux quanta. These additional fluxes attached to one electron make the composite fermions (CFs). There are even number of fluxes, as many as 8, attached to one electron.

   Let us consider the even flux quanta. Depending on the value of 
$n$,
let us say that $2pn\phi_o$=1 T, then for 2, 4, 6, and 8 quanta, 
the field becomes,
$$
B-2, \,\,\,  B-4, \,\,\,B-6, \,\,\, B-8,\,\,\, etc.\eqno(13)
$$
Consider B-2 = 12.5 T, then B=14.5 T and we expect features at B-4=10.5T, B-6= 8.5T, B-8= 6.5 T, etc. but {\sl there are no such 
features in the experimental data} at these fields. Instead of even feature (13)  there
 are features according to series (11). We have generated two series, one according to even number in (8) and the other according to (5)
 with charge determined by (11). The expression (8) for B$^*$ is independent of the effective charge $\nu$ or $\nu^*$. Therefore, 
the even features of (8) should be observed independent of the observation of the charge series.

\section{ Back Calculation}

     If we write 2 and $\pm 1$, then we can generate 1/3 and 1 charge. By using even numbers, all of the odd numbers can be generated.
Therefore, we obtain 2p$\pm$ 1, 4p$\pm$ 1, 6p$\pm$ 1, 8p$\pm$ 1, etc. 
Now, look for fields so that we can get $B\pm 2pn\phi_o$ and flux quantization gives, B=n$\phi_o$. The total field B$\pm2pn\phi_o$ can also be quantized so that B$^*$=n$\phi_o/\nu^*$. In this way we can generate the CF model. This means that odd denominators are observed
 in the charge but the construction made (8) is not correct. It 
violates the Biot and Savart's law. The series of fractions such as p/(2p$\pm$ 1) are
thus phenomenological and hence give the correct answers, which are based on the experimental data. However, in nature the field is 
not quantized as in (8).

\section {Conclusions}

 In conclusion, the simulation of the CF  model
 from the experimental data is not correct, i.e., it violates (a) 
the Biot and Savart's law, and (b) the even fields as in (13) are 
not observed. The composition of fields as in (8) with flux 
quantization in one term as in (8) is not correct. The eq.(12) is not consistent with the Biot and Savarts law. It has been found[5] that
 the CF are inconsistent 
with a fermion statistics. Apparently as Wilczek[6] first thought,
 the attachment of flux quanta to electrons is not yet found. Usually
when new discoveries are made they are consistent with the already 
known physics but  the CF is not.

\vskip1.0cm
\noindent{\bf References}
\begin{enumerate}
\item K. von Klitzing, G. Dorda and M. Pepper, Phys. Rev. Lett.
	{\bf45}, 494 (1980).
\item D.C. Tsui, H.L. St\"ormer and A.C. Gossard, Phys. Rev. Lett.
	{\bf48}, 1559 (1982).
\item W. Pan, H. L. Stormer, D. C. Tsui, L. N. Pfeiffer, K. W. Baldwin and K. W. West, Phys. Rev. Lett. {\bf88}, 176802(2002).
\item J.K. Jain, Phys. Rev. Lett. {\bf63}, 199 (1989).
\item K. N. Shrivastava, cond-mat/0202459.
\item F. Wilczek, Phys. Rev. Lett. {\bf48}, 957 (1982).

\end{enumerate}

\noindent Fig.1: {\sl Biot and Savart's law shows that (a) the current I
produces the field B and (b) a current of $I_{\phi}$ produces the field of 2pn$\phi_o$.(c) Two flux quanta attached to one electron and (d) four flux quanta attached to one electron. The attachment of flux quanta to one electron is not consistent with the Biot and Savart's law.}

\end{document}